\documentclass{article}

\usepackage{amsfonts}
\usepackage{amssymb}
\usepackage{epsfig}
\usepackage{psfig}

\usepackage{buckow}

\newcommand{\beqn}{\begin{eqnarray}}
\newcommand{\eeqn}{\end{eqnarray}}
\newcommand{\non}{\nonumber \\}

\def\titleline{\begin{flushright} \vspace{-2.5cm} 
{\small HU-EP-00/44 \\ \vspace{-0.2cm} 
hep-th/0010198} \end{flushright}
\vspace{1cm}
Magnetic Flux in Toroidal Type I Compactification}

\begin{document} 

\def\authors{
{Ralph Blumenhagen}\footnote{Email: blumenha@physik.hu-berlin.de}, 
{Lars G\"orlich}\footnote{Email: goerlich@physik.hu-berlin.de},
{Boris K\"ors}\footnote{Email: koers@physik.hu-berlin.de} 
and {Dieter L\"ust}\footnote{Email: luest@physik.hu-berlin.de} }

\def\addresses{Humboldt Universit\"at zu Berlin \\
Institut f\"ur Physik, Invalidenstr. 110, 10115 Berlin, Germany}

\def\abstracttext{
We discuss the compactification of type I strings on a torus with additional
background gauge flux on the D9-branes. The solutions to the cancellation
of the RR tadpoles display various phenomenologically attractive features:
supersymmetry breaking, chiral fermions and the opportunity to reduce the
rank of the gauge group as desired. We also point out the equivalence of the
concept of various different background fields and noncommutative deformations
of the geometry on the individual D9-branes by constructing 
the relevant boundary states to describe such objects. } 

\large
\makefront

\section{Introduction}

It is of conceptual interest on its own behalf to study the effects of
nontrivial background gauge fields in type I string theory
\cite{bac,bist,bgkl,aads}. In particular the
presence of constant magnetic flux on a torus 
still allows a microscopic description of
the relevant degrees of freedom in terms of the worldsheet theory. A very
prominent feature is that a constant
flux induces a deformation of the background geometry from a commutative towards
a noncommutative coordinate algebra \cite{schom,sw}. More precisely, we
discuss the presence 
of various different fluxes on different D9-branes, which gives rise to
noncommutative deformations of the coordinate algebras of open string
coordinates on the D9-branes, the deformation parameter governed by the
respective value of the flux. To prove the one-loop 
consistency of such compactifications we 
show that they still allow to cancel the RR tadpole of the Klein
bottle. Interestingly, D9-branes with nonvanishing flux also couple to tensor
fields of degree lower than ten, they carry D5-brane charge as well. \\

We find attractive phenomenological features for the effective theory of
type I compactified on a four- or sixdimensional torus with appropriate flux
on the D9-branes. The spectrum includes chiral fermions. The gauge group can
be engineered to be any product of unitary, orthogonal and symplectic groups
with the upper bound of 16 for the rank. Supersymmetry is always
broken, which implies that the one-loop amplitude is nonvanishing. A 
NSNS tadpole is left over and open string tachyons appear, leaving the
dynamical stability of the configuration an open question. \\

The paper is organized as follows. In section \ref{sec2} we first discuss
D-branes on a torus with an additional magnetic background field ${\cal F}$ 
on their world volume and its equivalence to a noncommutative deformation of
their open string 
coordinate algebra. We also show how a better intuition of such D-brane
set-ups can be obtained by employing a T-dual version with D-branes at
angles. In section \ref{sec3} 
we then compute the tadpole cancellation conditions for type I strings, when
the D9-branes carry such magnetic flux and discuss the implications of the
solutions. Finally we present a
semi-realistic example and point out two major obstructions to
obtain models with concrete phenomenological impact. 

\section{D-branes with ${\cal F}$-Flux, Non-commutativity and T-duality}
\label{sec2}

We compactify type I string theory on a six- or fourdimensional
torus, adding magnetic background flux on the D-branes. Therefore we first
perform a preliminary analysis of D-branes wrapping a simple twodimensional
torus $\mathbb{T}^2$, which carries a nontrivial $U(1)$ gauge field with
constant field strength on its world volume. The $\mathbb{T}^2$ is chosen to
have purely imaginary complex and K\"ahler structures 
for simplicity. They define the spectrum of bosonic zero-modes, KK momenta and
winding states. The constant background flux is given by the $U(1)$ gauge
curvature ${\cal F}_{ij}={\cal F} \delta_{ij}$. Physically inequivalent
background fields are classified by their first Chern number  
$m \in \mathbb{Z}\simeq H^2(\mathbb{T}^2,\mathbb{Z})$, which is 
the magnetic charge of the field configuration. The possible values for
${\cal F}$ are given by 
\beqn
{\cal F} = \frac{ 2\pi m}{n R_1R_2} ,
\eeqn 
where $n\in \mathbb{Z}$ denotes the electric charge quantum 
of the particular $U(1)$. 
Thus, any stack of coincident D$_\mu$-branes wrapping the torus is
charcterized by the two integers $n_\mu$ and $m_\mu$, the electric and
magnetic charge quanta of its worldvolume gauge theory. \\

The D-branes themselves are described in CFT by a corresponding
boundary state defined by the boundary conditions with
background ${\cal F}$-flux 
\beqn
\left( \partial_\sigma X_1 + {\cal F}_\mu\ \partial_\tau X_2 
\right) \vert_{\partial \Sigma_\mu} &=& 0 ,\non 
\left( \partial_\sigma X_2 - {\cal F}_\mu\ \partial_\tau X_1 
\right) \vert_{\partial \Sigma_\mu} &=& 0 . 
\eeqn
The mode expansion of the coordinate fields of a string stretching between two
D-branes $\mu$ and $\nu$ then displays Fourier modings in $\mathbb{Z}+\left
  ( \phi_\mu-\phi_\nu \right) /\pi$, where we defined $\phi_\mu \equiv {\rm
  arccot} ({\cal F}_\mu)$. It is by now well known and widely appreciated that
coordinates of open strings ending on D-branes with magnetic flux on their
world volume do not commute. The relevant ``open string metric'' and ``open
string antisymmetric tensor field'' are \cite{sw}
\beqn \label{noncmetric}
G_\mu^{ij} \equiv \frac{1}{1+{\cal F}_\mu^2} \delta^{ij} , \quad 
\theta_\mu^{ij} \equiv 
-\frac{2\pi{\cal F_\mu}}{1+{\cal F}_\mu^2} \epsilon^{ij}.
\eeqn
The commutator of the open string coordinates can then be expressed in terms
of the deformation parameter $\theta^{12}_\mu$
\beqn
\left[ X_1(\tau,\sigma) , X_2 (\tau,\sigma') \right] 
     \vert_{\sigma=\sigma'\in \partial \Sigma_\mu}= 
- \frac{2\pi {\cal F}_\mu}{1+{\cal F}_\mu^2} = i\theta_\mu^{12} .
\eeqn
The boundary state $\vert {\rm B}_\mu \rangle$ 
that solves the boundary conditions 
\beqn
\left( p_{\rm L} +p_{\rm R} -i{\cal F}_\mu \left( p_{\rm L} -p_{\rm R} \right)
\right) \vert 
{\rm B}_\mu \rangle &=& 0 , \non 
\left( \alpha_q + \exp \left( 2i\phi_\mu \right) \tilde{\alpha}_{-q} \right)
\vert {\rm B}_\mu \rangle &=& 0 
\eeqn
is represented by the coherent state (here the bosonic part)
\beqn \vert {\rm B}_\mu \rangle = 
    \exp \left( \sum_{q > 0}{\frac{1}{q}
    e^{2i\phi_\mu} \alpha_{-q}\tilde{\alpha}_{-q} } +\ {\rm c.c.}\right) 
    \sum_{r^1,r^2\in \mathbb{Z}}{\vert r^1,r^2 
    \rangle_{\left({\cal F}_\mu\right)}} .
\eeqn
The zero mode spectrum can be explicitly evaluated and it turns out that it
can be summarized by using (\ref{noncmetric}) as well:
\beqn
p_\mu^i = \frac{1}{1+{\cal F}^2_\mu} \frac{r^i}{n_\mu R_i} 
        = G_\mu^{ij} \frac{r^j}{n_\mu R_j} ,\quad
w_\mu^i = \frac{i{\cal F}_\mu \epsilon^{ij}}{1+{\cal F}^2_\mu}
            \frac{r^j}{n_\mu R_j}  
        = \frac{\theta_\mu^{ij}}{2\pi} \frac{r^j}{n_\mu R_j}  
\eeqn
giving rise to the open string mass spectrum
\beqn
M_{\mu,{\rm open}}^2 = \frac{r^i}{n_\mu R_i} G^{ij}_\mu \frac{r^j}{n_\mu R_j} .
\eeqn
We recognize the electric charge quantum $n_\mu$ entering as a winding number
for the D$_\mu$-brane. \\

Summarizing, the internal space of type I string
theory on a torus with extra magnetic fluxes on the D9-branes is 
noncommutative, while the noncompact gauge theory stays 
commutative. Furthermore, it is strictly equivalent to employ the open string
metric and antisymmetric tensor in all computations or to explicitly implement
the background gauge field into the boundary conditions. \\

There is another equivalence which we only note as an aside. The deformation
induced by the background gauge field is again identical to performing a 
left-right asymmetric rotation of the closed string coordinates \cite{asym}. 
This establishes an interesting link between asymmetric string vacua and
noncommutative geometry. Asymmetric nongeometric rotational 
symmetries are gauged 
in asymmetric orbifolds or orientifolds, such that these spaces do not
distinguish between certain values of the noncommutativity parameters
$\theta^{ij}$. In \cite{asym} a large class of asymmetric orientifolds with these
properties has been constructed, which are related to a previously explored
type of symmetric and therefore geometric orientifolds with D-branes at angels
\cite{angle1,angle2,angle3,bonn} 
via a certain T-duality, which we now come to discuss. \\

It is very helpful for the visualisation of D-branes with
various fluxes to perform a T-duality in one of the directions of the
$\mathbb{T}^2$, say $x_1$. The complex and K\"ahler structures 
get exchanged and the boundary conditions of open strings
on D$p$-branes with ${\cal F}_\mu$-flux switch to D$(p-1)$-branes at
an angle $\phi_\mu$ relative to the $x_1$ axis: 
\beqn
\partial_\sigma X_1 + {\cal F}_\mu\ \partial_\tau X_2 =0 
& \stackrel{T_1}{\longleftrightarrow} & 
\partial_\tau \left( X_1 + \cot(\phi_\mu) X_2 \right) =0 ,\non
 \partial_\sigma X_2 - {\cal F}_\mu\ \partial_\tau X_1 =0 
& \stackrel{T_1}{\longleftrightarrow} & 
\partial_\sigma \left( X_2 - \cot(\phi_\mu) X_1 \right) =0 .
\eeqn
This is illustrated in figure 1. 
\begin{figure}[h]
\label{fig1}
\begin{center}
\makebox[6cm]{
 \epsfxsize=11cm
 \epsfysize=3.5cm
 \epsfbox{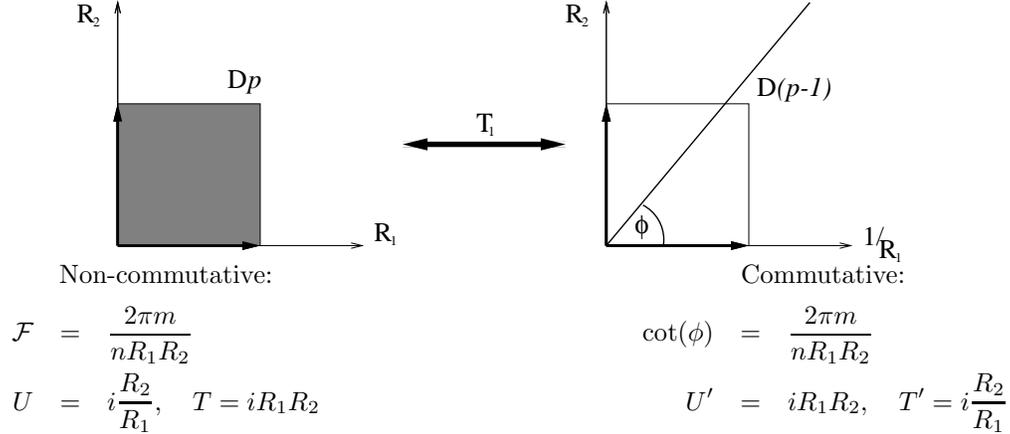}
}
\end{center}
\vspace{-.7cm}
\hspace{-4.5cm}
\hbox{
\vbox{
\begin{center}
Non-commutative:\\ \vspace{-.5cm}
\beqn
{\cal F} &=& \frac{2\pi m}{nR_1R_2} \non
U &=& i\frac{R_2}{R_1}, \quad T=iR_1R_2 
\nonumber \eeqn
\end{center}}
\hspace{-7.5cm}
\vbox{
\begin{center}
Commutative:\\ \vspace{-.5cm}
\beqn
\cot(\phi) &=& \frac{2\pi m}{nR_1R_2} \non
U'&=& iR_1R_2, \quad T'=i\frac{R_2}{R_1} 
\nonumber \eeqn
\end{center}}
}
\caption{T-duality of D-branes with flux} 
\end{figure}
%
The T-duality also identifies commutative and noncommutative internal
geometries. In the T-dual picture the electric and magnetic quantum numbers
that characterise the gauge theory on the D$p$-brane map to the winding
numbers of the D$(p-1)$-brane on the circles of the torus, $(n,m)\in
\mathbb{Z}^2\simeq H^1(\mathbb{T}^2,\mathbb{Z})$. 
%
In the following we shall always employ the T-dual ``branes at angles''
picture to illustrate the stacks of coincident D9-branes with different
magnetic fluxes, which we meet in the orientifold construction. \\

\section{Type I with ${\cal F}$-Flux}
\label{sec3}

Now we can proceed to the main object of this paper, the compactification of
type I strings on a torus with magnetic flux on the D9-branes. We choose the
$2d$-dimensional torus to be a product
$\mathbb{T}^{2d}=\mathbb{T}^2_{(j)}\times \cdots \times\mathbb{T}^2_{(j)}$ 
with each individual $\mathbb{T}^2_{(j)}$ 
of purely imaginary complex and K\"ahler
structure as before. The various 
radii we denote by $R_{1,2}^{(j)}$, the volumina by $V^{(j)}$, and the flux on
any individual stack of coincident D9$_\mu$-branes we call ${\cal
  F}^{(j)}_\mu$. The appropriate T-duality $T_1$ on all the $x_1$ directions
translates to a geometrical setting with D$(9-d)$-branes at angles
$\phi_\mu^{(j)}$. It also
affects the world sheet parity $\Omega$ which gets combined with a reflection
${\cal R}$ in all $x_1$ directions: $T_1 \Omega T_1^{-1} =  \Omega {\cal R}$. 
The standard procedure of the construction of type I compactifications as
orientifolds of type IIB next requires to compute the one-loop massless
tadpoles of the Klein bottle closed string amplitude and then introduce open
string sectors into the theory in order to cancel the divergencies.    
As the models we consider will be shown to break supersymmetry, the one-loop
amplitude does not vanish on the whole, in particular a NSNS tadpole
survives. Also there can be tachyons in the open 
string spectrum, their masses depending on the concrete choice
of the radii. Assuming some mechanism to stabilize their vacuum expectation
values, they may serve as Higgs bosons in the effective field theory. On the
other hand there are no closed string tachyons, which would signal a more
serious gravitational instability. \\

\subsection{One-loop amplitudes}

First note that the presence of ${\cal F}$-flux only affects the open string
sector, so that the Klein bottle amplitude remains unchanged. We include the
result for completeness: 
\beqn
{\cal K} &=& 2^{5-d} c\ (1-1) \int_0^\infty{
  \frac{dt}{t^{6-d}}\ 
  {\rm Tr} \left( 
  \frac{\Omega}{2} {\cal P}_{\rm GSO} 
  e^{-2\pi t \left( L_0 + \bar{L}_0 \right) } 
  \right) } \non
&=& 2^{3-d} c\ (1-1) \int_0^\infty{\frac{dt}{t^{6-d}} 
         \frac{\vartheta\left[ 0 \atop 1/2 \right]^4}{\eta^{12}} \prod_{j=1}^d
         \left( 
         \sum_{r,s\in\mathbb{Z}^2} e^{-\pi t \left( {r^2/ {R^{(j)}_1}^2 } +
              s^2 /{R^{(j)}_2}^2 \right) } \right) }
\eeqn
with $c= {V_{10-2d}}/\left( 8\pi^2 \right)^{5-d}$ and $\alpha'=1$. Its
contribution to 
the RR tadpole is to be found by a modular transformation into the tree
channel:  
\beqn
\tilde{\cal K} \sim \int_0^\infty{dl\ 2^{13-d} \prod_{j=1}^d {V^{(j)}} } .
\eeqn
We now add D9$_\mu$-branes into the background, at least some with
$\cot(\phi_\mu^{(j)})= {\cal F}^{(j)}_\mu \not= 0$, and compute the open string
diagrams. The symmetry under $\Omega$ also
forces to introduce the D9-branes pairwise, any $(n_\mu^{(j)},m_\mu^{(j)})$
accompanied by its image 
$(n_{\mu'}^{(j)},m_{\mu'}^{(j)})\equiv(n_\mu^{(j)},-m_\mu^{(j)})$. For 
two types $\mu$ and $\nu$ of D9-branes then at least four kinds of open
strings have to be regarded, which is depicted in figure 3: \\
\begin{figure}[h]
\label{fig3}
\begin{center}
\makebox[6cm]{
 \epsfxsize=6cm
 \epsfysize=5cm
 \epsfbox{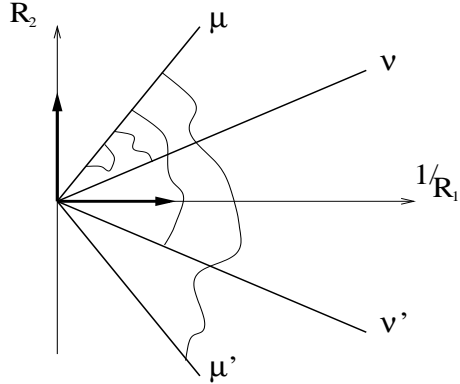}
}
\end{center}
\caption{Open string sectors}
\end{figure}
%

The entire annulus amplitude can then be decomposed into sectors as follows: 
\beqn
{\cal A} &=& c \int_0^\infty{\frac{dt}{t^{6-d}}\  {\rm Tr}_{\mbox{\scriptsize open}} 
\left( \frac{1}{2} {\cal P}_{\rm GSO} e^{-2\pi tL_0} 
\right) } \non
&=& \sum_\mu{  
        \left( {\cal A}_{\mu\mu} +
         {\cal A}_{\mu\mu'}+  \left( \mu \leftrightarrow \mu' \right) \right)
       } + \non 
& & \sum_{\mu<\nu}{\left( {\cal A}_{\mu\nu} +{\cal A}_{\mu\nu'} + 
         {\cal A}_{\mu'\nu}+ {\cal A}_{\mu'\nu'} + \left( \mu, \mu'
           \leftrightarrow \nu, \nu' \right) \right) } ,
\eeqn
where each term has to be weighted by the intersection number $I_{\mu \nu}$ of
the two branes in question. By general topological arguments this is 
\beqn
I_{\mu \nu} =  \prod_{j=1}^d{ \left( m^{(j)}_\mu n^{(j)}_\nu - n^{(j)}_\mu
    m^{(j)}_\nu \right) } .
\eeqn
The result finally reads
\beqn
{\cal A}_{\mu \nu} &=& c\ 2^{-2}\ {\rm N}_\mu {\rm N}_\nu I_{\mu\nu}
      \int_0^\infty{\frac{dt}{t^{6-d}} 
      \sum_{\alpha,\beta\in\{0,1/2\}} (-1)^{2(\alpha+\beta)}
            e^{2 i \alpha\sum_j{(\phi^{(j)}_\nu-\phi^{(j)}_\mu)}}
            e^{i\pi d/2}} \non
& & \hspace{5cm} \times\ \frac{\vartheta\left[-\beta \atop \alpha\right]^{4-d} 
   \prod_{j=1}^{d}{\vartheta\left[-(\phi^{(j)}_\nu-\phi^{(j)}_\mu)/
        \pi-\beta \atop \alpha\right]}}{\eta^{12-3d}
      \prod_{j=1}^{d}{\vartheta\left[-(\phi^{(j)}_\nu-\phi^{(j)}_\mu)/\pi-1/2
          \atop 1/2\right] }} 
\eeqn
leading to the following contribution to the massless RR tadpole
\beqn
\tilde{\cal A}_{\mu\nu} 
\sim \int_0^\infty{dl\ 2^{3-d}\ {\rm N}_\mu {\rm N}_\nu\ 
  \prod_{j=1}^d{\left( \frac{m^{(j)}_\mu m^{(j)}_\nu}{V^{(j)}} + 
                n^{(j)}_\mu n^{(j)}_\nu V^{(j)}  \right) }} .
\eeqn
Interestingly, the existence of terms proportional to the volume of some
the $\mathbb{T}^2$, signals that D9-branes with
additional gauge flux not only carry the charge of a conventional D9-brane but
also charges of D$(9-2i)$-branes, $i=1...d$. Further, we again meet the
electric quantum numbers $n_\mu$ rescaling the charges: A D9-brane wrapping
some $\mathbb{T}^2$ twice just carries twice as much 
charge. This fact will be shown to enable a very simple method to lower the
rank of the resulting gauge group. \\

The final contribution comes from the M\"obius strip, resulting only from open
strings invariant under $\Omega$, i.e. stretching between some D9$_\mu$-brane
and its image D9$_{\mu'}$-brane:
\beqn
{\cal M}= c\ \int_0^\infty{\frac{dt}{t^{6-d}}\  {\rm Tr}_{\mbox{\scriptsize open}} 
\left( \frac{\Omega}{2} {\cal P}_{\rm GSO} e^{-2\pi tL_0} \right)
} =\sum_\mu{\left( {\cal M}_{\mu \mu'} + {\cal
      M}_{\mu' \mu} \right) } .
\eeqn
Now the individual sectors have to be weighted with the number $I_{\mu
  \mu'}^{(\Omega)}$ of invariant intersections only. The result is
\beqn 
{\cal M}_{\mu \mu'} = c\ 2^5\ {\rm N}_\mu I_{\mu
    \mu'}^{(\Omega)} (-1)^{d} \int_0^\infty{dl\  
\sum_{\alpha,\beta\in\{0,1/2\}}{(-1)^{2(\alpha+\beta)}
      e^{4i \alpha\sum_j{\phi^{(j)}_\mu}}
            \frac{\vartheta
    \left[-\beta \atop \alpha \right]^{4-d} \prod_{j=1}^{d}{\vartheta\left
        [ -2\phi^{(j)}_\mu/ \pi-\beta \atop \alpha \right] }}{\eta^{12-3d}
    \prod_{j=1}^{d}{\vartheta\left[ -2\phi^{(j)}_\mu/\pi-1/2 \atop 1/2
      \right]}}}} 
\nonumber \eeqn
and its contribution to the massless RR tadpole: 
\beqn 
\tilde{\cal M}_{\mu \mu'} \sim \int_0^\infty{dl\ 2^{9-d}\ {\rm N}_\mu\   
           \prod_{j=1}^d{n^{(j)}_\mu V^{(j)} }} .
\eeqn 
The amplitudes combine into the perfect square and the cancellation in, for
instance, six dimensions ($d=2$) requires:
\beqn
& & {\mbox{D9-brane charge,}}\quad V^{(1)}V^{(2)}:
\sum_\mu{{\rm N}_\mu n_\mu^{(1)} n_\mu^{(2)}} = 16 ,
\non  
& & {\mbox{D5-brane charge,}}\quad  \frac{1}{V^{(1)}V^{(2)}}:
\sum_\mu{{\rm N}_\mu m_\mu^{(1)} m_\mu^{(2)}} =0  .
\eeqn
The charges which would correspond to D7-branes cancel by the $\Omega$
symmetry of the D-brane setting.  
In four dimensions with $d=3$ all three kinds of D5-brane charges, identified
by one volume factor in the numerator and two in the denominator, have to
be cancelled, while the D7- and D3-brane charges again vanish automatically. 
Pure D9-branes with vanishing magnetic flux enter the tadpole
cancellation with $m_\mu=0$ and D5-branes corresponding to infinite flux with
$n_\mu=0$. We have also calculated the conditions for a tadpole cancellation
in the $\mathbb{Z}_2$ orbifold, where in addition to the D9-brane charge as
well a net D5-brane charge of 16 is required. This opens additional options
for the construction of interesting models, such as vacua entirely without
D5-branes. It even appears to be possible to maintain supersymmetry 
\cite{aads}. \\

\subsection{Gauge group, chirality and supersymmetry}

As a D9-brane with ${\cal F}\not=0$ is not invariant under $\Omega$, but
maps to the brane of opposite flux, there is no $\Omega$ projection 
in its open string spectrum. The resulting gauge group on a stack of N
such branes is therefore $U({\rm N})$ instead of $SO({\rm N})$ or $Sp({\rm
  N})$. One can then directly realize gauge groups of the type 
\beqn
SO({\rm N}_9) \times Sp({\rm N}_5) \times \prod_\mu{U({\rm N}_\mu)} ,
\eeqn
and has also got the possibility to lower the rank simply by
choosing any ``electric'' winding number $n^{(j)}_\mu >1$. 
The fourdimensional 
spectrum of massless fermions is generic, independent of any radii, and
chiral in any open string sector, where two D9-branes have
${\cal F}^{(j)}_\mu - {\cal F}^{(j)}_\nu \not=0$ for all $j$. 
The multiplicities of states are essentially given by the intersection
numbers of the respective D6-branes in the T-dual picture and the general results
for the spectrum have been presented in \cite{bgkl}.  
Supersymmetry is actually always broken, even if sometimes in a rather subtle
fashion. 

\subsection{An example}

\begin{figure}[h]
\label{fig4}
\begin{center}
\makebox[6cm]{
 \epsfxsize=12cm
 \epsfysize=7cm
 \epsfbox{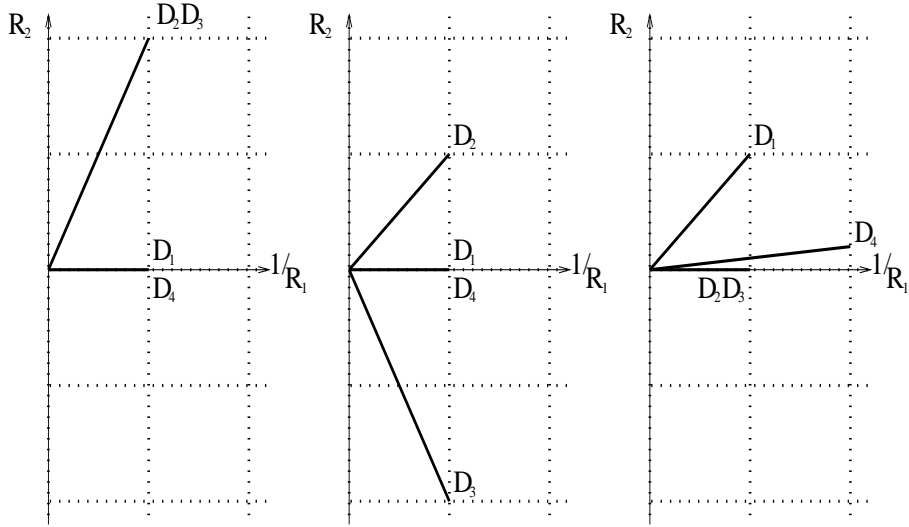}
}
\end{center}
\caption{The four generation D-brane setting} 
\end{figure}

In this final section we shall briefly mention an example which is meant to
demonstrate the power and 
simplicity of a ``bottom-up'' stategy of model construction,
which consists in only 
putting fluxes on type I D9-branes to engineer the desired gauge
group and spectrum. The D-brane configuration shown in figure \ref{fig4} \\
%

supports a gauge group $U(3) \times U(2) \times U(1)^2$ in the effective
  theory. One of the four abelian factors is found to be anomalous, which is
  consistent with a single Green-Schwarz mechanism to be possible in order to
  cancel the anomaly and decouple the gauge boson. 
The fermion spectrum includes four generations of standard model fermions with
  right handed neutrinos included, and one of the nonanomalous $U(1)$ factors
  has suitable quantum numbers to serve as hypercharge. \\

Unfortunately, two major drawbacks need to be mentioned which prevent a
really phenomenological model building so far: 
First, the number of fermion generations is always even, which has its reason
in the arithmetics of intersection numbers. But even more troublesome, a
``large volume'' compactification is not compatible with a chiral spectrum. 
Any two D-branes with
unequal flux on all $\mathbb{T}^2$ leave no transverse direction that could
become large. Again this can be better understood in the dual picture, where
any two D$(9-2d)$-branes at nonvanishing relative angles always span the entire
internal space. Therefore the string scale cannot be chosen at the electroweak
scale and supersymmetry breaking is obsolete. \\[.5cm]

\noindent
{\large \bf Acknowledgements} \\

\noindent
The content of this article has been presented by R.B. at the Ninth Marcel
Grossmann Meeting at Rome, Italy, in July 2000, and by B.K. at the RTN Network
conference at Berlin, Germany, in Octobre 2000. The work was partly supported
by the EEC contract ERBFMRXCT96-0045, the DFG 
and by the Studienstiftung des deutschen Volkes.

\end{document}